\documentclass[12pt]{article}

\newcommand{\bbr}{I\!\! R}
\newcommand{\bbz}{Z\!\!\! Z}

\newcommand{\2}{$^2$}
\newcommand{\3}{$^3$}
\newcommand{\4}{$_4$}
\newcommand{\5}{$_5$}
\newcommand{\x}{arXiv:}

\usepackage{graphicx}

\begin{document}
\thispagestyle{empty}
\begin{center}

\null \vskip-1truecm \vskip2truecm {\bf APS Instability and the
Topology of the Brane-World\\} \vskip1truecm Brett McInnes
\vskip1truecm

 National University of Singapore

email: matmcinn@nus.edu.sg\\

\end{center}
\vskip1truecm \centerline{ABSTRACT} \baselineskip=15pt
\medskip

As is well known, classical General Relativity does not constrain
the topology of the spatial sections of our Universe. However, the
Brane-World approach to cosmology might be expected to do so,
since in general any modification of the topology of the brane
must be reflected in some modification of that of the bulk.
Assuming the truth of the Adams-Polchinski-Silverstein conjecture
on the instability of non-supersymmetric AdS orbifolds, evidence
for which has recently been accumulating, we argue that indeed
many possible topologies for accelerating universes can be ruled
out because they lead to non-perturbative instabilities.
PACS-1996: 98.80.Cq, 11.25.-w Keywords: Branes, Topology, AdS
Orbifolds

\addtocounter{section}{1}
\section*{1. A Theoretical Perspective on Cosmic Topology }
It is clear that Einstein's equation alone does not fix the
topology of a cosmological model \cite{kn:levin}. Since there are
many possible topologies consistent with the familiar FRW
geometries, it is natural to ask: what physical principle
\emph{does} fix topology in cosmology? Here we consider this
question in the light of the observed acceleration of the Universe
\cite{kn:carroll}, which may indicate that the basic
\emph{geometry} [though not necessarily the global
\emph{topology}] of our world is that of de Sitter spacetime.

The widely popular brane-world approach \cite{kn:maartens} to
cosmology allows us to attack this problem. For when de Sitter
spacetime is formulated in this way
\cite{kn:zerbini}\cite{kn:branenewworld}, as a brane-world in
AdS\5, the conformal infinity of the brane-world actually resides
on the conformal infinity of AdS\5. [See \cite{kn:orbifold} for
the details.] It follows that, in the brane-world picture,
non-trivial spatial topology in cosmology necessarily implies
non-trivial topology for the boundary of the \emph{local} AdS\5 in
which the brane is embedded. This in turn gives us a possible way
of testing the physical acceptability of candidate topologies,
since the physics of AdS\5 and its orbifolds
\cite{kn:orbifold}\cite{kn:ross}\cite{kn:ofar} has been studied
intensively. The boundary of the standard simply-connected version
of AdS\5 has topology $\bbr \; \times \;$S\3, so a locally de
Sitter brane with non-trivial topology will have to be embedded in
a version of AdS\5 which has a boundary where S\3 is replaced by
some non-singular quotient. Taking such quotients will certainly
affect the AdS/CFT dual field theory, for example in the way
recently discussed by Dowker \cite{kn:dowker}. [It would also have
major effects in the context of the proposed ``dS/CFT
correspondence"
\cite{kn:strominger}\cite{kn:mcnees1}\cite{kn:mcnees2}, which
however we shall not be using here.] Again, it turns out that such
\emph{non-singular} boundary quotients give rise to orbifold
singularities in the bulk. It is this effect that we shall study
here.

In AdS\5 there are sources of instability which arise when one
considers \emph{non-supersymmetric} orbifolds. This was pointed
out by Adams, Polchinski, and Silverstein \cite{kn:adams}, who
conjectured that the condensation of closed string tachyons coming
from the twisted sector would tend to resolve the orbifold
singularity and restore supersymmetry. This restoration of the
``deficit angle" cannot, however, be confined to the vicinity of
the (former) singularity: the jump in the deficit is produced by a
dilaton pulse which expands outward at the speed of light,
ultimately restoring the geometry to its pre-orbifold state.
Strong evidence in favour of this conjecture has recently been
obtained by studying both the late-time structure
\cite{kn:gregory}\cite{kn:headrick} and the internal consistency
of the proposed mechanism \cite{kn:moore}\cite{kn:zwiebach}. It
has been argued by Horowitz and Jacobson \cite{kn:jacob} that a
similar phenomenon can be expected in non-supersymmetric orbifolds
of AdS. The AdS/CFT correspondence then predicts a similarly
radical instability for the matter fields on the de Sitter brane.
The upshot is that the \emph{brane-world picture must be
considered inconsistent if the brane-world is required to reside
in an} AdS\5 \emph{orbifold which is not supersymmetric}. Since
topologically non-trivial de Sitter branes are associated with
AdS\5 orbifolds, we clearly have here a potentially powerful
criterion for ruling out many candidate topologies: we must check
whether the relevant AdS\5 orbifold is supersymmetric.

There are infinitely many purely spatial quotients of dS\4. For
the sake of clarity we shall concentrate on one of these, namely
the de Sitter version of the ``dodecahedral Universe" proposed in
\cite{kn:weeks2}. The general case then follows by similar
techniques. [We focus on this particular topology because it
illuminates the general case. We stress that we have nothing to
say here about the motivation or observational status of the
dodecahedral model: for that, see \cite{kn:cornish}.]

We begin with a brief explanation of the structure of the
dodecahedral space in the context of de Sitter cosmology. We then
examine the corresponding AdS\5 orbifold and show explicitly that
it has no surviving supersymmetries. In view of the above, we can
use this to \emph{rule out} the dodecahedral topology, and,
similarly, many other candidate topologies, assuming the validity
of the Adams-Polchinski-Silverstein argument.

\addtocounter{section}{1}
\section*{2. The Dodecahedral Cosmos as a Brane-World}

The de Sitter solution of the Einstein equation is valid for
\emph{any} three-manifold having the local geometry of S\3.
However, even if we confine ourselves to ``Copernican" models,
that is, those with spatial sections which are homogeneous, then
there are still infinitely many \emph{locally} spherical
candidates to be considered. These fall into an ADE classification
of the kind familiar to string theorists: there are two infinite
families together with a special class consisting of just three
(isometry classes of) manifolds. The most complex of these,
corresponding to E$_8$ in the ADE classification, is the
Poincar\'e dodecahedral space, also known as the Poincar\'e
homology sphere. It is obtained simply by identifying all of the
opposite faces of a dodecahedron, after consistently applying a
$\pi$/5 twist. (The other two spaces in the E-series are obtained
in an analogous way from the regular tetrahedron and the regular
octahedron.) One can obtain a basic model of an accelerating
Universe in this way by replacing the S\3 spatial sections of de
Sitter spacetime with copies of the Poincar\'e dodecahedral space,
thereby giving the dodecahedral Universe the basic dynamics of an
accelerating spacetime.

Topologically, the dodecahedral space has the structure
S\3/\~I$_{120}$, where \~I$_{120}$ is a finite subgroup of SU(2).
This group is called the binary icosahedral group; it is a group
of 120 elements, such that \~I$_{120}$/$\bbz_2$ = I$_{60}$, the
icosahedral group. This is the 60-element group of symmetries of a
regular dodecahedron or icosahedron, the dual polyhedron of the
dodecahedron. (Throughout this work, ``symmetries" of a polygon or
polyhedron will always mean ``orientation-preserving symmetries in
three dimensions".) Since I$_{60}$ is a group of symmetries of a
geometric object (it is a subgroup of SO(3)), it is easier to
visualise than \~I$_{120}$, and this will be useful to us.

Combining these observations, we can obtain an accelerating
Universe with the Poincar\'e dodecahedral space as spatial
sections simply by taking de Sitter spacetime dS\4(S\3) and factoring S\3 by \~I$_{120}$, to obtain
dS\4(S\3/\~I$_{120}$). If we do this, we obtain a spacetime which
is \emph{locally} indistinguishable from de Sitter spacetime, but
which has a different global structure. In particular, while
dS\4(S\3) is spatially homogeneous and globally isotropic,
dS\4(S\3/\~I$_{120}$) is homogeneous but not globally isotropic.

Now let us embed this version of de Sitter spacetime in the
appropriate version of AdS\5. Five-dimensional anti-de Sitter
spacetime, AdS\5, is defined as the locus
\begin{equation}\label{eq:A}
- A^2 - B^2 + w^2 + x^2 + y^2 + z^2 = -L^2,
\end{equation}
in a flat six-dimensional space of signature (2,4). This is a
space of constant negative curvature $-$1/L\2. It is not hard to
see that in AdS\5 there is a copy of dS\4 at each point of the
bulk which is sufficiently ``near" to the boundary. To be precise,
there is such a copy corresponding to each value of B such that
$|$B$|$ $>$ L. Choosing coordinates on AdS\5 which cover this
region \emph{only}, one can in fact \cite{kn:orbifold} express the
AdS\5 metric as
\begin{eqnarray}\label{eq:C}
g(AdS_5) = d\rho^2 + \textup{sinh}^2(\rho /L)\;[ -d\tau^2 + L^2
\textup{cosh}^2 (\tau /L)\{d\chi^2 +
\textup{sin}^2(\chi)[d\theta^2 +
\textup{sin}^2(\theta)d\phi^2]\}],
\end{eqnarray}
or
\begin{equation} \label{eq:D}
g(AdS_5) =  d\rho^2 + \textup{sinh}^2(\rho/L)g(dS_4),
\end{equation}
where $g$(dS$_4$) is the usual global metric for de Sitter
spacetime. Thus, we can put a de Sitter brane at $\rho$ = c for
some constant c; points in AdS\5 corresponding to larger values of
$\rho$ are cut away, in the usual Randall-Sundrum manner. However,
the time coordinate on the brane is related to the global radial
AdS\5 coordinate r by the equation
\begin{equation}\label{eq:E}
\textup{sinh}(r/L) = \;\textup{sinh}(c/L)\;\textup{cosh}(\tau /L),
\end{equation}
so we see that the \emph{temporal} conformal infinity of the brane
($\tau \rightarrow \pm\infty$) actually resides on the
\emph{spatial} conformal infinity of the bulk (r $\rightarrow
\infty$). Thus the brane still has access to the conformal
infinity of the bulk, despite the cutting away of the region $\rho
>$ c. It follows that if we factor S\3 in the de Sitter brane by a
finite group such as \~I$_{120}$, \emph{then we have no option but
to do the same} to the S\3 in the boundary of AdS\5. That is, we
are forced to allow \~I$_{120}$ to act on the coordinates w, x, y,
and z in equation (\ref{eq:A}) and then take the quotient. We can
do this because \~I$_{120}$ is contained in the isometry group of
AdS\5; in fact it just acts on the angular coordinates in equation
(\ref{eq:C}), preserving the spherical part of the metric.

Recall now that the spatial sections of AdS\5 are copies of the
hyperbolic space H$^4$. \emph{Any} finite group of isometries of
H$^4$ has a (common) fixed point, and so, unlike
dS\4(S\3/\~I$_{120}$), the quotient AdS\5/\~I$_{120}$ is singular:
it is an orbifold. One might suspect that this orbifold
singularity at the centre of AdS\5 arises from the special, highly
symmetric geometry of AdS\5, but \emph{this is not correct}: no matter how 
we perturb the geometry of the quotient, it remains singular unless (perhaps)
the perturbation is so large that some curvature becomes positive. This follows from
a theorem of Cartan (\cite{kn:kobayashi}, page 111); see
\cite{kn:orbifold} for the details. This means that we still
expect an AdS\5 orbifold to be the correct background here even if
the exact geometry near the origin is not identical to that of
AdS\5.

Thus, if the dodecahedral model is valid, then this tells us that
the bulk is an orbifold. The symmetry group of this AdS\5 orbifold is
given by
\begin{equation}\label{eq:EE}
\textup{Isom}(AdS_5/\tilde{I}_{120})\; = \;\textup{O}(2) \;
\times\; \textup{SO(3)};
\end{equation}
this agrees with the conformal group of the quotient
CCM$_4$/\~I$_{120}$, where CCM$_4$ is the conformal
compactification of Minkowski space; this is of course in accord
with AdS/CFT expectations. (Note that when AdS\5 is obtained as a
string background, orientation-reversing isometries are not matter
symmetries, so in this context we should state the symmetry group
as SO(2)$\;\times\;$SO(3) rather than O(2)$\;\times\;$SO(3).) We
see that factoring by finite groups drastically reduces the size
of the spacetime isometry group, from fifteen dimensions to four,
from non-compact to compact. This prepares us for the still more
drastic reduction of supersymmetry to be discussed below.

\addtocounter{section}{1}
\section*{3. Stringy Instability of AdS\5/\~I$_{120}$}

Quotients of \emph{flat} spacetimes by ADE finite groups have been
studied extensively; see for example \cite{kn:myers}. The survival
of supersymmetry in such cases can often be understood in terms of
holonomy theory. In particular, taking the quotient of $\bbr^4$ by
a finite subgroup of one of the SU(2) factors of SO(4) results in
an orbifold with holonomy large enough to break half of the
supersymmetries.

The case of orbifolds of AdS\5 is quite different. For whereas
$\bbr^4$ has trivial holonomy, AdS\5 already has the maximal
possible holonomy group for a (time and space orientable)
Lorentzian five-manifold, namely SO$^+$(1,4). Since the action of
\~I$_{120}$ on AdS\5 preserves time and space orientation (that
is, the action does not involve time, and the Poincar\'e
dodecahedral space is orientable in the ordinary sense, since
\~I$_{120}$ is completely contained in SO(4), not just O(4) ), it
follows that taking the quotient of AdS\5 by \~I$_{120}$ cannot
change the holonomy group in any way: it is already as large as it
can be if no orientation is reversed. Hence we cannot extend our
intuitions regarding the preservation of supersymmetry from the
flat case to the anti-de Sitter case. Fortunately, the question of
supersymmetry on finite group quotients of anti-de Sitter space
has been studied \cite{kn:mukhi}, and the degree to which
AdS\5/\~I$_{120}$ is supersymmetric can be settled by means of an
explicit calculation.

First, let us simplify the problem as follows. Inspection of the
regular dodecahedron reveals that its symmetry group, I$_{60}$,
contains the symmetry group of the tetrahedron, T$_{12}$. (There
is a standard way to fit a tetrahedron inside a dodecahedron; see
http://www.divideo.it/personal/todesco/java/polyhedra/dodecahedron\_tetrahedron.html
for an excellent picture of this. Ignore the symmetries of order 5
associated with the pentagonal faces. The remaining symmetries are
just those which define T$_{12}$. In the same way one sees that
T$_{12}$ is a subgroup of the group, O$_{24}$, of symmetries of  a
regular octahedron.) The tetrahedral group has only 12 elements.
Inspection of the regular tetrahedron reveals that T$_{12}$ in its
turn contains a (normal) subgroup isomorphic to $\bbz_2 \times
\bbz_2$. (Each $\bbz_2$ is generated by a symmetry of the
tetrahedron which acts by rotation through $\pi$ about an axis
joining the midpoints of a chosen pair of opposite edges. There
are three such pairs of opposite edges, but a combination of the
two rotations corresponding to any two pairs generates the
rotation corresponding to the third, so the group consists of two
copies of $\bbz_2$, not three. The obvious $\bbz_3$ symmetry of
the tetrahedron permutes the three non-trivial elements of $\bbz_2
\times \bbz_2$.) Thus T$_{12}$, and therefore I$_{60}$, contain
$\bbz_2 \times \bbz_2$ in a natural way. When we lift I$_{60}$ to
\~I$_{120}$, we must therefore also lift $\bbz_2 \times \bbz_2$ to
a subgroup of SU(2), and it is not hard to show that this subgroup
is Q$_8$, the quaternionic group \{$\pm$1, $\pm$i, $\pm$j,
$\pm$k\}, where i, j, and k are the usual basis quaternions; here
we are thinking of SU(2) as the group of all unit quaternions, the
symplectic group Sp(1). (One sees that Q$_8$ projects to $\bbz_2
\times \bbz_2$ by pretending that i, j, and k commute and square
to +1 instead of $-$1.) Thus Q$_8$ is contained in \~T$_{24}$, the
binary tetrahedral group; since, as we saw above, T$_{12}$ is a
subgroup of both O$_{24}$ and I$_{60}$, it follows that Q$_8$ is
also contained in the binary octahedral group \~O$_{48}$ and also,
most importantly, in the binary icosahedral group \~I$_{120}$.

Now AdS\5 can be represented using quaternions by taking the
coordinates used in equation (\ref{eq:A}) and defining
\begin{eqnarray}\label{eq:F}
D    & = &  A + iB                  \nonumber \\
C    & = &  w + ix + jy + kz.                   \nonumber \\
\end{eqnarray}
If $\widehat{C}$ represents the quaternion conjugate of C, defined
by reversing the sign of the vector part of the quaternion but not
its scalar part, then the definition of AdS\5 may be written as
\begin{equation}\label{eq:G}
-\widehat{D}D + \; \widehat{C}C = - L^2.
\end{equation}
We see at once from this that Q$_8$ acts on AdS\5 by q : (D, C)
$\rightarrow$ (D, qC) for each q $\in$ Q$_8$, since $\widehat{qC}
= \widehat{C}\widehat{q}$ and $\widehat{q}q$ = 1. As Q$_8$ is
generated by i and j, the action of Q$_8$ on AdS\5 can be fully
understood by studying the effect of these two elements. Since we
have
\begin{eqnarray}\label{eq:GG}
i(w + ix + jy + kz)    & = &  -x + iw - jz + ky                  \nonumber \\
j(w + ix + jy + kz)    & = &  -y + iz + jw - kx,                   \nonumber \\
\end{eqnarray}
the action of Q$_8$ on AdS\5 is therefore fully described by the
maps
\begin{eqnarray}\label{eq:H}
i     & : & (A, B, w, x, y, z) \rightarrow  (A, B, -x, w, -z, y)                   \nonumber \\
j     & : & (A, B, w, x, y, z) \rightarrow  (A, B, -y, z, w, -x),                   \nonumber \\
\end{eqnarray}
where we denote the map by the corresponding quaternion.

In order to make a comparison with the work of Ghosh and Mukhi
\cite{kn:mukhi}, let us switch from quaternions to ordinary
complex coordinates for the embedding space of AdS\5, with Z$_i$,
i = 1,2,3, defined by
\begin{eqnarray}\label{eq:I}
Z_1 & = & A + iB                    \nonumber \\
Z_2 & = & w + ix                       \nonumber \\
Z_3 & = & y + iz,                      \nonumber \\
\end{eqnarray}
so that AdS\5 is
\begin{equation}\label{eq:J}
-Z_1 \overline{Z_1} + Z_2 \overline{Z_2} + Z_3\overline{Z_3} = -
L^2,
\end{equation}
where the bar denotes the ordinary complex conjugate. A useful set
of coordinates ($\theta_1,\theta_2,\delta,\alpha,\beta$) is
defined \cite{kn:mukhi} by
\begin{eqnarray}\label{eq:K}
Z_1 & = & L\textup{cosh}({{\theta_1}\over{2}})e^{i\delta}                   \nonumber \\
Z_2 & = & L\textup{sinh}({{\theta_1}\over{2}})\textup{cos}({{\theta_2}\over{2}})e^{i\alpha}                   \nonumber \\
Z_3 & = & L\textup{sinh}({{\theta_1}\over{2}})\textup{sin}({{\theta_2}\over{2}})e^{i\beta},                      \nonumber \\
\end{eqnarray}
and the Killing spinors on AdS\5 are given by \cite{kn:mukhi}
\begin{equation}\label{eq:L}
\epsilon =
e^{{{1}\over{4}}\Gamma_4\theta_1}e^{-{{1}\over{4}}\Gamma_{14}\theta_2}e^{-{{1}\over{2}}\Gamma_{24}\alpha}
e^{{{1}\over{2}}\Gamma_3\delta}e^{{{1}\over{2}}\Gamma_{15}\beta}\epsilon_0,
\end{equation}
where the $\Gamma_i$ all square to unity except for $\Gamma_3$
(which squares to $-$1) and where $\epsilon_0$ is a constant
spinor.

Now in terms of the Z$_i$ coordinates, the action of i and j given
in equations (\ref{eq:H}) are expressed as
\begin{eqnarray}\label{eq:M}
i     & : & (Z_1, Z_2, Z_3) \rightarrow  (Z_1, iZ_2, iZ_3)                  \nonumber \\
j     & : & (Z_1, Z_2, Z_3) \rightarrow  (Z_1, - \overline{Z_3},  \overline{Z_2});                   \nonumber \\
\end{eqnarray}
notice that \emph{both} of these square to the map $(Z_1, Z_2,
Z_3) \rightarrow  (Z_1, -Z_2, -Z_3)$, and they anti-commute, as
they should according to the quaternion multiplication table. In
terms of the coordinates given by equations (\ref{eq:K}), the
actions of i and j are given by
\begin{eqnarray}\label{eq:N}
i     & : & (\theta_1,\theta_2,\delta,\alpha,\beta) \rightarrow  (\theta_1,\theta_2,\delta,\alpha+{{\pi} \over {2}},\beta+{{\pi} \over {2}})                  \nonumber \\
j     & : & (\theta_1,\theta_2,\delta,\alpha,\beta) \rightarrow  (\theta_1,\theta_2+\pi,\delta,-\beta,-\alpha).                   \nonumber \\
\end{eqnarray}
We can now see the effects of i and j on the Killing spinor
$\epsilon$ given by equation (\ref{eq:L}):
\begin{eqnarray}\label{eq:O}
i     & : & \epsilon \rightarrow
e^{{{1}\over{4}}\Gamma_4\theta_1}e^{-{{1}\over{4}}\Gamma_{14}\theta_2}e^{-{{1}\over{2}}\Gamma_{24}(\alpha+{{\pi}\over{2}})}e^{{{1}\over{2}}\Gamma_3\delta}e^{{{1}\over{2}}\Gamma_{15}(\beta+{{\pi}\over{2}})}\epsilon_0
                   \nonumber \\
j     & : & \epsilon \rightarrow  e^{{{1}\over{4}}\Gamma_4\theta_1}e^{-{{1}\over{4}}\Gamma_{14}(\theta_2+\pi)}e^{{{1}\over{2}}\Gamma_{24}\beta}e^{{{1}\over{2}}\Gamma_3\delta}e^{-{{1}\over{2}}\Gamma_{15}\alpha}\epsilon_0.                   \nonumber \\
\end{eqnarray}
Now suppose that we construct the quotient AdS\5/Q$_8$, an
orbifold which contains a non-singular brane-world with the local
geometry of de Sitter spacetime but with S\3/Q$_8$ as spatial
sections. (Note that S\3/Q$_8$ can be visualised by simply taking
a cube and identifying all opposite faces after a consistent
rotation by $\pi/2$.)  Then this quotient will retain some
supersymmetry if $\epsilon$ is invariant with respect to
\emph{both} i and j. From the first equation in the set
(\ref{eq:O}), we see at once that for $\epsilon$ to be invariant
with respect to i, the constant spinor $\epsilon_0$ has to satisfy
\begin{equation}\label{eq:P}
\Gamma_{24}\epsilon_0 = \Gamma_{15}\epsilon_0.
\end{equation}
Of course, not every $\epsilon_0$ can satisfy this, but some do:
in fact \cite{kn:mukhi}, there is a two-dimensional space of
solutions of (\ref{eq:P}), and so the quotient AdS\5/$\bbz_4$,
where $\bbz_4$ is generated by i, retains precisely half of the
supersymmetries. Similarly, the quotient of AdS\5 by the $\bbz_4$
generated by j is also half-supersymmetric. But now suppose that
we require $\epsilon$ to be invariant with respect to \emph{both}
i and j. Then, noting that neither i nor j affects $\theta_1$, we
see that the condition for the invariance of $\epsilon$ under the
action of j is
\begin{equation}\label{eq:PP}
e^{-{{\pi}\over{4}}\Gamma_{14}}e^{{{1}\over{2}}\Gamma_{24}\beta}e^{{{1}\over{2}}\Gamma_3\delta}e^{-{{1}\over{2}}\Gamma_{15}\alpha}\epsilon_0
=
e^{-{{1}\over{2}}\Gamma_{24}\alpha}e^{{{1}\over{2}}\Gamma_3\delta}e^{{{1}\over{2}}\Gamma_{15}\beta}\epsilon_0.
\end{equation}
But now, \emph{using equation (\ref{eq:P})} --- that is, requiring
simultaneous invariance under i and j --- we can define a spinor
$\eta$ by
\begin{equation}\label{eq:PPP}
\eta =
e^{{{1}\over{2}}\Gamma_{24}\beta}e^{{{1}\over{2}}\Gamma_3\delta}e^{-{{1}\over{2}}\Gamma_{15}\alpha}\epsilon_0
=
e^{-{{1}\over{2}}\Gamma_{24}\alpha}e^{{{1}\over{2}}\Gamma_3\delta}e^{{{1}\over{2}}\Gamma_{15}\beta}\epsilon_0,
\end{equation}
and then equation (\ref{eq:PP}) becomes simply
\begin{equation}\label{eq:Q}
e^{-{{\pi}\over{4}}\Gamma_{14}}\eta = \eta,
\end{equation}
but this is not possible except for trivial $\epsilon_0$. Thus
some supersymmetry generators can survive factoring by
\emph{either} i or j --- \emph{but none can survive both}.

We conclude that AdS\5/Q$_8$ is a non-supersymmetric orbifold of
AdS\5. (That it is indeed an orbifold and not a manifold can be
seen from equations (\ref{eq:H}): clearly all those points of the
form (A,B,0,0,0,0), with A\2 + B\2 = L\2 (see equation
(\ref{eq:A})) are left unmoved by every element of Q$_8$.) But we
saw earlier, using the geometry of the regular polyhedra, that
Q$_8$ is a subgroup of all of the binary polyhedral groups. Since
no Killing spinor on AdS\5 can survive factoring by Q$_8$, it
follows that \emph{no Killing spinor is invariant by those groups
either}, and we see that all of the spaces AdS\5/\~T$_{24}$,
AdS\5/\~O$_{48}$, and AdS\5/\~I$_{120}$ are
\emph{non-supersymmetric orbifolds. }

In fact, of all the homogeneous quotients of S\3, the only ones
that lead to a supersymmetric quotient of AdS\5 are those in the
A-series of the ADE classification mentioned above in section 2.
To see this, note that we have already dealt with the three
E-groups, \~T$_{24}$, \~O$_{48}$, and \~I$_{120}$, so we can turn
to the D-groups and then the A-groups. The D-groups are the
\emph{generalized quaternionic groups}, Q$_{4n}$, of order 4n, for
all n $\geq$ 2. For n $\geq$ 3 they are the groups which cover the
\emph{dihedral groups}, D$_{2n}$, the groups of symmetries of the
regular n-sided polygons; that is, Q$_{4n}$/$\bbz_2$ = D$_{2n}$.
We can regard Q$_{4n}$ as being generated by the quaternion i
together with another unit quaternion q of order 2n. A somewhat
more intricate version of the calculation given above shows that,
as in the case of Q$_8$, there are Killing spinors which can
survive factoring by the cyclic groups generated by \emph{either}
i or q, but none can survive factoring by both. (This actually
follows from our discussion above if n is even, for then q can be
chosen to be a root of j, but a separate argument is needed when n
is odd.) Thus none of the quotients AdS\5/Q$_{4n}$ is
supersymmetric.

Next, the ``A-quotients" are the (homogeneous) lens spaces,
generalizing the quotient by either i or j but not both. It is
clear that all of these lead to quotients of AdS\5 which
\emph{are} supersymmetric: they are half-supersymmetric, since the
quotients (by cyclic groups of any order) are like the quotients
of AdS\5 by the $\bbz_4$ generated by i \emph{or} j, which retain
a two-dimensional space of Killing spinors.

Finally we note that there is a huge class of S\3 quotients
\cite{kn:weeks3} which are not homogeneous; these are usually
ignored for ``Copernican" reasons, though one can question whether
we have the right to assume that we are not at a special place in
space, given that we \emph{do} seem to find ourselves at a special
point in time, a time when the dark energy has ``recently" begun
to dominate \cite{kn:carroll}. ``de Sitter" spacetimes with the
simplest inhomogeneous lens spaces as spatial sections are
obtained as brane-worlds in an AdS\5 orbifold --- recall that the
action by \emph{any} finite group on the spatial sections has a
fixed point --- by factoring AdS\5 by the $\bbz_m$ generated by
the map
\begin{equation}\label{eq:R}
(Z_1,Z_2,Z_3) \rightarrow  (Z_1,\gamma Z_2,\gamma^bZ_3),
\end{equation}
where $\gamma$ is a primitive m\emph{th} root of unity and b is an
integer, relatively prime to m, with 1 $<$ b $\leq$ m/2. For a
Killing spinor to survive this projection, condition (\ref{eq:P})
above is replaced by
\begin{equation}\label{eq:S}
\Gamma_{24}\epsilon_0 = b\Gamma_{15}\epsilon_0.
\end{equation}
However, the eigenvalues of the matrix $-\Gamma_{24} +
b\Gamma_{15}$ can easily be computed \cite{kn:mukhi}: they are
\begin{equation}\label{eq:T}
(1 + b), -(1 + b), (1 - b), -(1 - b).
\end{equation}
In view of the conditions on b, none of these is zero, and so
(\ref{eq:S}) cannot be satisfied by any non-trivial $\epsilon_0$.
This proves that de Sitter branes with inhomogeneous lens spaces
as spatial sections cannot reside in a supersymmetric AdS\5
orbifold. Since the other inhomogeneous quotients of S\3 are all
obtained \cite{kn:weeks3} by factoring by groups which contain
subgroups acting, after extension from the brane to AdS\5, as in
(\ref{eq:R}), we see that none of the versions of de Sitter
spacetime with inhomogeneous spatial sections can occur as
brane-worlds in supersymmetric AdS\5 orbifolds. All of these
results can be verified tediously but explicitly by noting that
all elements of SO(4), including those which act on S\3 such that
the quotient is not homogeneous, can be represented by a pair of
unit quaternions (q$_1$, q$_2$), modulo $\pm$(1, 1), acting on a
quaternion C by C $\rightarrow$ q$_1$Cq$_2^{-1}$. If C is the
quaternion given in equation (\ref{eq:F}), then in the coordinates
given by (\ref{eq:K}) we have
\begin{equation}\label{eq:U}
C =
L\;\textup{sinh}({{\theta_1}\over{2}})[\textup{cos}({{\theta_2}\over{2}})\textup{cos}(\alpha)
+ i\;\textup{cos}({{\theta_2}\over{2}})\textup{sin}(\alpha) +
j\;\textup{sin}({{\theta_2}\over{2}})\textup{cos}(\beta) +
k\;\textup{sin}({{\theta_2}\over{2}})\textup{sin}(\beta)],
\end{equation}
and it is therefore possible to compute explicitly the action of
any element of SO(4) on the Killing spinor in equation
(\ref{eq:L}) by means of quaternion multiplication. The results
agree with those obtained above.

We have seen explicitly that AdS\5/\~I$_{120}$ is a
non-supersymmetric orbifold. In fact, we have a much stronger
statement. Combining all of the results of the present section, we
see that among all of the possible actions by finite groups on
S\3, only a small subset extend from the brane to AdS\5 in such a
way that the quotient is supersymmetric. This subset consists of
actions by finite cyclic groups such that the quotient
S\3/$\bbz_n$ is homogeneous: that is, the S\3 quotient is a
homogeneous lens space. The final conclusion is that among all the
versions of de Sitter spacetime with topologically non-trivial
spatial sections, the only ones which can be self-consistently
interpreted as brane-worlds \emph{within string theory} are the
ones with homogeneous lens spaces as spatial sections. (In
addition, there are other ways of modifying the topology of de
Sitter spacetime, involving quotients which affect the time axis.
Most of these can be ruled out in the same way: see
\cite{kn:orbifold}.)

\addtocounter{section}{1}
\section*{4. Conclusion}

The idea that the spatial sections of the four-dimensional
Universe should take the form S\3/[non-trivial finite group] is
extremely natural from the string point of view. For such
constructions have arisen before: the famed Calabi-Yau manifolds
used in compactifications of heterotic E$_8 \;\times$ E$_8$ string
theory are precisely of the form [compact Riemannian
manifold]/[non-trivial finite group], the non-triviality being
necessary for gauge symmetry breaking by   ``Wilson loops" (see
\cite{kn:kachru} for a recent discussion of this). Among the vast
variety of quotients of S\3, the dodecahedral space
S\3/\~I$_{120}$ has a strong claim to be the most interesting;
among many other remarkable properties, it corresponds to E$_8$ in
the ADE classification of the homogeneous quotients of S\3. It is
remarkable that it cannot arise as a model for the spatial
sections of an accelerating brane-world cosmology in string
theory. In fact, the only survivors of APS instability are the
homogeneous lens spaces, which clearly deserve further study.

\end{document}